\documentstyle[epsfig,12pt,aasms4]{article}
\begin{document}

\lefthead{Fender et al.}
\righthead{Cir X-1}
\slugcomment{Submitted to ApJ Letters}

\title{An asymmetric arcsecond radio jet from Circinus X-1}

\author{Robert Fender\altaffilmark{1}, Ralph Spencer\altaffilmark{2},
Tasso Tzioumis\altaffilmark{3}, Kinwah Wu\altaffilmark{4}, \\
Michiel van der Klis\altaffilmark{1}, Jan van Paradijs\altaffilmark{1,5}, Helen
Johnston\altaffilmark{6}} \altaffiltext{1}{Astronomical Institute
`Anton Pannekoek', University of Amsterdam, and Center for High Energy
Astrophysics, Kruislaan 403, 1098 SJ Amsterdam, The Netherlands}
\altaffiltext{2}{University of Manchester, Nuffield Radio Astronomy
Laboratories, Jodrell Bank, Macclesfield, Cheshire SK 11 9DL, U.K.  }
\altaffiltext{3}{Australia Telescope National Facility, CSIRO, Paul
Wild Observatory, Narrabri NSW 2390, Australia}
\altaffiltext{4}{Research Centre for Theoretical Astrophysics, School
of Physics, University of Sydney, NSW 2006, Australia }
\altaffiltext{5}{Physics Department, University of Alabama in
Huntsville, Huntsville AL 35899, USA}
\altaffiltext{6}{Anglo-Australian Observatory, P.O. Box 296, Epping
NSW 2121, Australia}

\begin{abstract}

In observations with the Australia Telescope Compact Array we have
resolved the radio counterpart of the unusual X-ray binary Cir X-1
into an asymmetric extended structure on arcsecond scales.  In order
to quantify the asymmetry we have redetermined as accurately as
possible both the optical and radio coordinates of the source. The
extended emission can be understood as a compact, absorbed core at the
location of the X-ray binary, and extended emission up to 2 arcsec to
the southeast of the core. The arcsec-scale extended emission aligns
with the larger, more symmetric arcmin-scale collimated structures in
the surrounding synchrotron nebula.  This suggests that the transport
of mass and/or energy from the X-ray binary to the synchroton nebula
is occurring via the arcsec-scale structures.  The ratio of extended
flux from the southeast to that from the northwest of the core is at
least 2:1. Interpreted as relativistic aberration of an intrinsically
symmetric jet from the source, this implies a minimum outflow velocity
of 0.1 c. Alternatively, the emission may be intrinsically asymmetric,
perhaps as a result of the high space velocity of the system.

\end{abstract}


\section{Introduction}

Circinus X-1 is a highly variable, at times very luminous X-ray
binary, with an orbital period of 16.6 days, as revealed by the
periodic recurrence of sudden drops or enhancements in its X-ray flux
and radio flaring (Kaluzenski et al. 1976; Haynes et al. 1978) . The
system is optically identified with the most reddened of three faint
stars lying within 1.5 arscec of each other in sky (Moneti 1992;
Duncan, Stewart \& Haynes 1993; see also HST image in Fig 1).  Stewart
et al. (1993) have reassessed published distance estimates to Cir X-1
and concluded that the most likely distance to the system is $\sim
6.5$ kpc.

For some time Cir X-1 was considered as a black-hole candidate because
of its rapid erratic X-ray flux variations (see Dower et al. 1982 for
a review of the older literature on Cir X-1), but the detection of
type I X-ray bursts, caused by thermonuclear flashes on the surface of
the neutron star (Tennant, Fabian \& Shafer 1986a, Tennant, Fabian \&
Shafer 1986b) established that Cir X-1 is an accreting neutron star.
In the 1970s the periodic radio flares of Cir X-1 peaked at $>
1$ Jy; since then its radio flux has decreased dramatically, and in
the past 10 yr it has rarely been detected at a level above 50 mJy
(e.g. Stewart et al. 1991). The source is embedded within a
synchrotron nebula and may be associated with the nearby supernova
remnant G321.9--0.3. Within the synchrotron nebula observations at 6.3
cm with an angular resolution of 12 arcsec have revealed collimated
structures which appear to be swept back towards G321.9--0.3 (Stewart et
al. 1993).

\section{Observations}

We have imaged Cir X-1 at high angular resolution ($\sim 1$ arcsec at
3.5 cm) with the Australia Telescope Compact Array (ATCA), on 1995
Sept 2, 1996 June 1, 1998 Feb 5 and Feb 23.  Fig 1 shows the image of
Cir X-1 from the 1998 Feb 23 observations, which had the best
observing conditions, revealing a strong ($\geq 30 \sigma$) extension
to the southeast.  The image was formed from 12 hr of observations at
3.5 cm with baselines between 214 and 5970 m. Observations were
regularly (2.5 of every 21.5 min) interleaved with those of a nearby
phase calibrator PMN J1524--5903 and of a compact source 6 arcmin to
the north, which we designate J1520.6--571.  Absolute flux calibration
was achieved with reference to PKS B1934--638; the data were reduced
using the MIRIAD software (Sault, Teuben \& Wright 1995).  The
observations were carefully planned to occur around binary orbital
phase $\sim 0.5$, and inspection of the X-ray light curve at the time
(as monitored with the Rossi XTE {\em All Sky Monitor}; Levine et
al. 1996) reveal that the observations occurred at least three days
after the end of the previous periodic outburst. No radio flaring
occurred during the 12 hr observation.  Mapping of the nearby compact
source J1520.6--571 (Fig 2), with the same phase calibration, produces
an almost-perfect point source (residuals from model point subtraction
$\leq 1$\%) ruling out phase errors as an origin for the extension of
Cir X-1.  The extension, though less pronounced, is also evident in
simultaneous observations with the same array at 6.3 cm.  An extension
to the southeast is also present in our earlier observations. The
arcminute-scale nebula and structures around the source (Stewart et
al. 1993) are undetected due to resolution effects, implying that they
contain no bright components on angular scales of arcseconds or less.

It is clear from Fig 1 that the extended emission is more intense to
the southeast of the emission peak than in other directions, but in
order to quantify this asymmetry it is necessary to determine as
accurately as possible the location of the X-ray binary itself (and
hence, presumably, the point of origin of any jets). We have proceeded to
do this in two ways, by redetermining as accurately as possible the
optical coordinates for Cir X-1 and by fitting simple models to the
radio map.

We have determined a new position for the optical counterpart to Cir
X-1 by performing astrometry on the red Second Epoch Sky Survey plate
(original glass negative) for field 177 from the UK Schmidt
Telescope. Positions of stars near our object were measured from
SuperCOSMOS scans of the plate; the astrometric solution was
determined using stars from the Tycho-ACT catalogue. We then tied this
astrometric solution to stars in a WFPC-1 archival image of the field
taken by HST (see Fig 1), using a linear fit between the coordinates
determined by SuperCOSMOS and the rectangular coordinates on the WFPC
mosaic image. This fit was used to derive the position for the optical
counterpart of Cir X-1.  Our improved optical position is RA
15:20:40.87, Dec --57:10:00.18 (J2000.0) with an error of $\pm 0.3$
arcsec in each coordinate.

The simplest adequate model for the radio map of Cir X-1 presented in
Fig 1 is a simultaneous fit of a 2D Gaussian and a point source
(fitting of a point source only without a simultaneous Gaussian
strongly over-subtracts at the emission peak). This fit is illustrated
in Fig 3, which is a flux profile along the jet axis (as indicated by
the lines in Fig 1). The same model fit was applied to the 6.3 cm
data, and flux densities for the total structure, and the point and
Gaussian components are listed in Table 1. The positive and negative
spectral indices for the Point and Gaussian components, respectively,
agree with expectations of an inverted spectrum, self-absorbed core
and steep spectrum, optically thin jets as seen in other X-ray
binaries.  Confidently associating the point source with unresolved
emission close to the binary, we derive improved radio coordinates for the
X-ray binary of RA 15:20:40.84, Dec --57:10:00.48
(J2000) with an error of up to 0.25 arcsec in both RA and Dec, arising
from uncertainties in the exact coordinates of the phase reference
source PMN J1524--5903. This is compatible within combined
uncertainties with the optical position derived above (radio to
optical offset is $\sim 0.4$ arcsec to the northwest, illustrated by
the overlay of the radio contours on the HST image in Fig 1).

The emission is reasonably well fit by the Gaussian component
described above and illustrated in Fig 3.  This component is clearly
offset to the southeast of the X-ray binary coordinates stated
above. Given the limitations imposed by the modelling, we can
conservatively place a lower limit of 2:1 on the ratio of fluxes in
the southeast to northwest jets. The lower limit on the ratio is very
similar if using instead the optical coordinates (as the location
along the asymmetry axis is almost identical).  We cannot at this time
exclude significantly higher ratios.

\section{Discussion}

There are two distinct probable origins for the observed asymmetric
structure, an intrinsically symmetric jet which appears one-sided
because of relativistic aberration, or a structure which for some
reason is intrinsically asymmetric. We briefly discuss both
possibilities below. 

\subsection{A relativistic jet}

The asymmetry in the extended radio emission is suggestive
of relativistic aberration of an intrinsically symmetric twin jet
inclined at some angle to the line of sight. This effect is well
observed in the relativistic jets of the X-ray binary GRS 1915+105
(Mirabel \& Rodriguez 1994), the core of SS 433 (Vermeulen et
al. 1993); and possibly Cygnus X-3 (Mioduszewski et al. 1998).  In
this case the stronger emission to the southeast would be the
approaching jet.  We note that the jet aligns extremely well with the
large scale structures reported by Stewart et al. (1993) shown by the
lines drawn on Fig 1, indicating the position angle of those
structures which we estimate to be $110 \pm 10$ degrees.  This
suggests that whether or not the jet is relativistic, it is
the route by which mass and/or energy is flowing from the X-ray binary
to the larger structures in the surrounding synchrotron nebula.  The
ratio of approaching to receding jet flux, for an intrinsically
symmetric jet, is given by

\[
\frac{S_a}{S_r} = \left( \frac{1+\beta \cos \theta}{1-\beta \cos \theta}
\right)^{k-\alpha}
\]

\noindent where $\alpha$ is the spectral index of the radio emission ($S_{\nu}
\propto \nu^{\alpha}$), $\beta$ is the intrinsic bulk velocity of the
jet as a fraction of the speed of light, $\theta$ is the angle of the
jet to the line of sight, and $k=2$ for a continuous jet and $k=3$ for
discrete ejections. From table 1 a spectral index of -1.2 is used for
the extended component. As $(S_a / S_r) \geq 2$, we derive

\[
\beta \cos \theta \geq 0.13
\] 

\noindent for a continuous jet, and 

\[
\beta \cos \theta \geq 0.09
\]

\noindent for discrete components. Thus the minimum velocity for an
intrinsically symmetric radio jet in Cir X-1 (for $\theta=0$ and
discrete components) is 0.1c. We note that our inferred velocities are
in conflict with the limit of $\leq 0.1$c on expansion velocity from an
early southern hemisphere VLBI observation, derived by Preston et
al. (1983), apparently confirmed by further VLBI observations by
Preston et al. (1989). However, we believe that the failure to resolve
the source following an outburst may have resulted from the compact
structure on milliarcsecond scales being dominated by the Doppler
boosted approaching component, in which case the limit applies to the
expansion of the component itself and not the outflow velocity.

Deceleration of jets on arcsec scales could affect the conclusions we
draw from arcsec images about the jets in Cir X-1. However, previous
experience has shown that this is not a serious problem.  Hjellming \&
Johnston (1981) determined the proper motions and structure of the SS
433 radio jet on an angular scale of up to 5 arcsec. The physical
properties derived by those VLA observations were later shown to be
entirely compatible with much higher resolution VLBI mapping
(e.g. Vermeulen et al. 1993). Furthermore the superluminal radio jets
of GRS 1915+105, as observed by Mirabel \& Rodriguez (1994) with the
VLA displayed a flux asymmetry on angular scales $\geq$ 1 arcsec which
is in agreement within a factor of order unity with flux asymmetries
observed later on much smaller scales with MERLIN (Fender et al.,
1998), and which are a direct result of relativistic aberration. In
fact there is no convincing evidence for the deceleration of ejected
radio components from X-ray binaries on angular scales up to several
arcseconds. The flux ratios of separate ejections of very different
magnitudes from GRS 1915+105 are all comparable (Mirabel \& Rodriguez
1994; Fender et al. 1998), implying the same jet formation and
acceleration process is in action for ejections of very different
sizes. However, as discussed below, larger radio structures associated
with some X-ray binaries are more symmetric than their compact jets,
suggesting that deceleration of ejecta does take place on angular
scales between several arcsec and arcmin.

\subsection{Intriniscally asymmetric emission}

The apparent one-sidedness of the extended emission may be intrinsic.
While an intrinsically one-sided jet is possible, this is not favoured
theoretically (e.g. Wiita 1991) and it seems instead likely that any
asymmetry is determined by the environment beyond the point of
origin. However, this is not seen in other X-ray binaries, where large
radio structures which may not be moving at relativistic velocities
are observed to be roughly symmetric. For example, symmetric radio
lobes of extent $\sim 6$ arcsec were reported around the X-ray binary
Cyg X-3 by Strom, van Paradijs \& van der Klis (1989) and have been
recently confirmed by Peracaula, Paredes \& Mart\'\i\
(1998). 

Perhaps the proposed high space velocity of Cir X-1 (a `runaway' X-ray
binary -- e.g. Stewart et al. 1993) is responsible for the observed
asymmetry. If the system has a high space velocity, a jet may be
continually having to propagate into a dense, inhomogenous medium. If
the binary really is moving north away from G321.9--0.3, then
clearly the southeast jet would feel less resistance from the ISM than
its northwest counterpart. However, we would not expect this effect to
be significant unless the space velocity of the binary and the outflow
velocity were of the same order of magnitude, which would rule out a
relativistic jet, and explanation of the larger collimated structure
to the NW then becomes problematic.

\section{Conclusions}

We have presented radio observations of Cir X-1 with an angular
resolution of $\sim 1$ arcsec which clearly resolve the radio
counterpart of the system into a core and extended emission to the
southeast at a position angle of $\sim 100$ degrees and an angular
extent of $\sim 2$ arcsec. In order to quantify the asymmetry we have
redetermined as accurately as possible the radio and optical coordinates
for the source. We find these to be in agreement within uncertainties
and to imply an asymmetry in the fluxes of extended emission to the
southeast and northwest of at least 2:1.

We have interpreted this asymmetric emission in two ways. Firstly we
discuss the apparent flux asymmetry which would be produced by a
relativistic jet inclined at some angle to the line of
sight. Following these arguments, and assuming an intrinsically
symmetric jet and ballistic motions of ejecta we derive a minimum
outflow velocity of 0.1 c. In this case the much more symmetric
collimated radio structures on arcmin scales implies the deceleration
of the ejecta on angular scales of tens of arcseconds. As the compact
object in Cir X-1 is a neutron star this may be the first
observational evidence that the formation of relativistic jets does
not require the presence of an event horizon or some other unique
property of black holes.  We have also discussed the possibility that
the extended radio emission is intrinsically asymmetric, possibly due
to the large space velocity of the system. In either case we think it
is likely that the extended emission on arcsec scales
from Cir X-1 originates in compact jets from the X-ray binary.

\acknowledgements
RPF thanks Jim Caswell and George Nicolson for several useful
discussions.  The Australia Telescope is funded by the
Commonwealth of Australia for operation as a National Facility managed
by CSIRO.  We are grateful to the SuperCOSMOS Unit at the ROE for
providing the scanned data from the UKST Southern sky red survey, and
to Nigel Hambly for providing assistance with the data.  The optical
image is based on observations made with the NASA/ESA Hubble Space
Telescope, obtained from the data archive at the Space Telescope
Science Institute. RPF was funded during the period of this research
by ASTRON grant 781-76-017 and EC Marie Curie Fellowship ERBFMBICT 972436.

\clearpage

\figcaption{The asymmetric radio jet in Cir X-1. (a) Top panel. 
Enhanced section of Fig 2 of Stewart et al. (1993) showing extended
structures on arcminute-scales extending SE and NW from Cir X-1. (b) 
Lower panel. Overlay of our new high-resolution radio image at
3.5 cm from 1998 Feb 23 (contours) over a high-resolution optical
image from the HST WFPC-1 (greyscale). This entire panel is
contained within the single central resolution element of the upper
panel. The peaks of radio and optical emission
correspond to the new, more accurate coordinates we have derived (see
text).  Contours are at -5, 5, 7.5, 10,
15, 20, 30, 50, 75 and 100 times the r.m.s. noise of 40 $\mu$Jy per
beam. The solid ellipse in the top left hand corner 
represents the synthesised beam, 1.31 $\times$ 1.17 arcsec at a
p.a. of -4.0 degrees. Section from Stewart et al. (1993) reproduced
by permission of Blackwell Scientific Publishers from MNRAS 1993, 261,
pp 593-598.}

\figcaption{Image at same frequency, from same observing run, using
same phase calibration as for Cir X-1 image in Fig 1, of a nearby
compact source which we designate J1520.6-571 (see text).  Contours
are at the same multiples of the r.m.s. noise, in this case 70 $\mu$Jy
per beam.  The beam is 1.34 $\times$ 1.17 arcsec at a p.a. of -14.6
degrees.  Subtraction of a point source from this image leaves
residuals of $\leq 1$\% of the total flux.  The lack of any
significant emission beyond a point source profile completely rules
out an origin for the extension of Cir X-1 in phase errors.}

\figcaption{A profile of the flux density of the image of Cir X-1
presented in Fig 1 along the jet axis (as indicated  by
lines in Fig 1), integrated over a strip of width two arcsec.
Negative offset indicates a southeasterly direction. Also
indicated are the results of a simultaneous fit of a point source and
Gaussian, and residuals from this fit (see text). No residuals are
greater than 8\% at any point.  Assuming the point source is
associated with the core of the X-ray binary, the asymmetry in the
structure of $\geq$ 2:1 from the SE to the NW must reflect Doppler
boosting of the approaching (SE) side of the jet or an intrinsic
asymmetry.}

\clearpage
\begin{figure*}
\centerline{
\epsfig{file=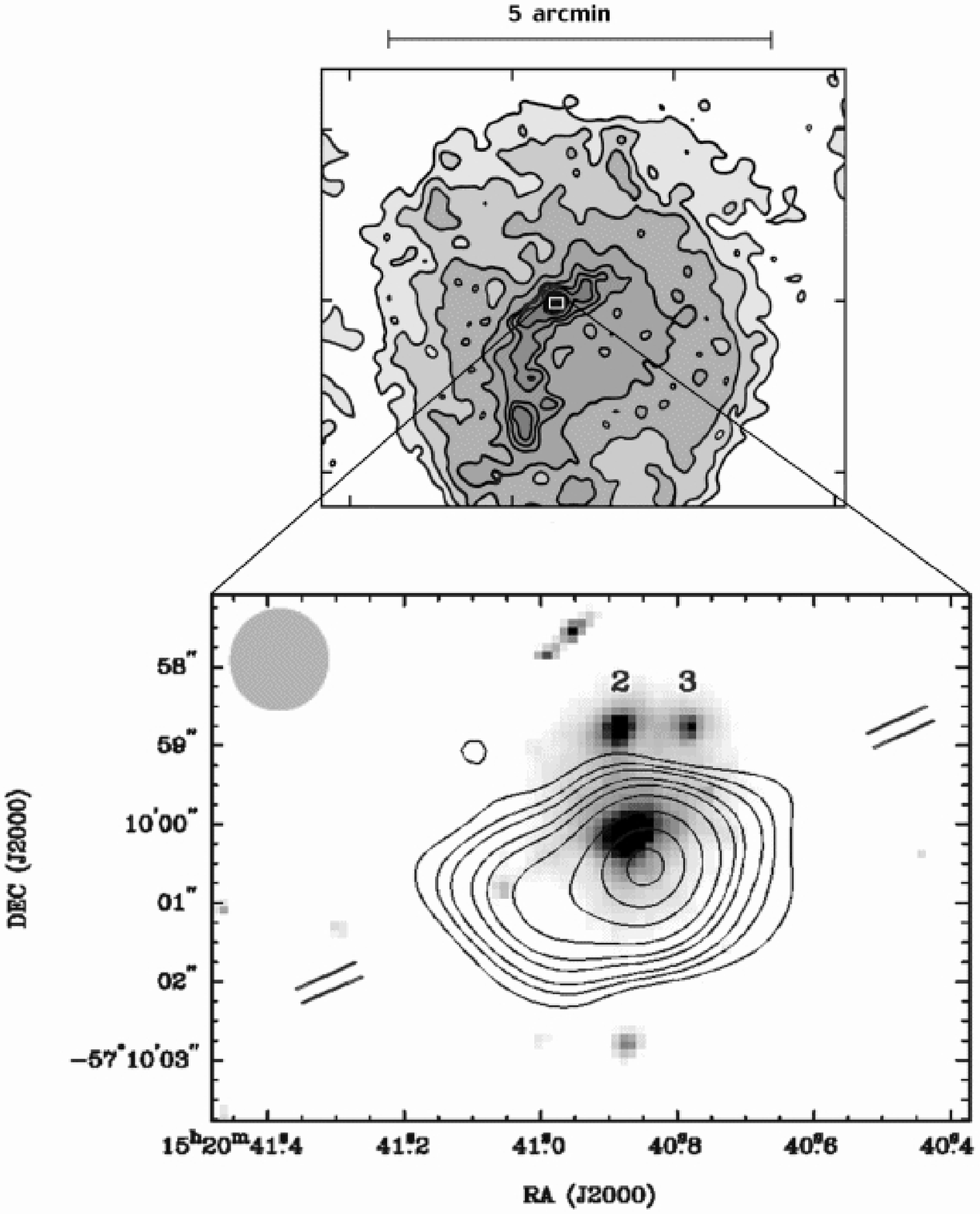,angle=0,width=14cm,clip}
}
\end{figure*}

\clearpage
\begin{figure*}
\centerline{
\epsfig{file=j1520.ps,angle=270,width=14cm,clip}
}
\end{figure*}

\clearpage

\begin{figure*}
\centerline{
\epsfig{file=cut.ps,angle=270,width=12cm,clip}
}
\end{figure*}

\clearpage

\begin{deluxetable}{|c|ccc|}
\tablecaption{Measured flux densities at 6.3 \& 3.5 cm for Cir X-1 from
1998 Feb 23 data. Total flux (i.e. inclusive of point and extended
structure, measured directly from maps within 5$\sigma$ contours) as
well as fluxes of simultaneous Point plus Gaussian fits are tabulated.
Overall spectral index corresponds to optically thin emission, but 
resolution into extended and point-like components reveals core to
have an inverted (self-absorbed) spectrum, supporting its association
with the compact core of the binary.}
\startdata
Component & $S_{\rm 6.3cm}$ (mJy) & $S_{\rm 3.5cm}$ (mJy) & $\alpha$ \nl
\tableline
Total (5$\sigma$ contours) & 9.9 & 7.7 & -0.7 \nl
Gaussian (model fit)       & 8.5 & 4.3 & -1.2 \nl
Point (model fit)          & 1.7 & 2.8 & +0.8 \nl
\enddata
\end{deluxetable}

\end{document}